\documentclass[aps,prb,reprint,superscriptaddress,notitlepage]{revtex4-1}
\usepackage{times,amsmath,amsfonts,amssymb,mathrsfs,graphics,graphicx,color,comment,bm}
\usepackage[next]{inputenc}
\usepackage[dvips]{epsfig}
\usepackage[pdfstartview=FitH]{hyperref}
\usepackage{natbib}
\usepackage{pdfsync}
\usepackage{appendix}
\usepackage{textcomp}

\begin{document}

\title{Measuring the Formation Energy Barrier of Skyrmions in Zinc Substituted Cu$_2$OSeO$_3$}

\author{M. N. Wilson}
\address{Durham University, Department of Physics, South Road, Durham, DH1 3LE, United Kingdom}
\author{M. Crisanti}
\address{Institut Laue-Langevin, Large Scale Structures Group, 71 avenue des Martyrs CS 20156, 38042, Grenoble, Cedex 9, France}
\address{University of Warwick, Department of Physics, Coventry, CV4 7AL, United Kingdom}
\author{C. Barker}
\address{Durham University, Department of Physics, South Road, Durham, DH1 3LE, United Kingdom}
\author{A. {\v S}tefan{\v c}i{\v c}}
\address{University of Warwick, Department of Physics, Coventry, CV4 7AL, United Kingdom}
\author{J.S. White}
\address{Laboratory for Neutron Scattering and Imaging (LNS), Paul Scherrer Institut (PSI), CH-5232 Villigen PSI, Switzerland}
\author{M. T. Birch}
\address{Durham University, Department of Physics, South Road, Durham, DH1 3LE, United Kingdom}
\address{Diamond Light Source, Didcot, OX11 0DE, United Kingdom}
\author{G. Balakrishnan}
\address{University of Warwick, Department of Physics, Coventry, CV4 7AL, United Kingdom}
\author{R. Cubitt}
\address{Institut Laue-Langevin, Large Scale Structures Group, 71 avenue des Martyrs CS 20156, 38042, Grenoble, Cedex 9, France}
\author{P .D. Hatton}
\address{Durham University, Department of Physics, South Road, Durham, DH1 3LE, United Kingdom}

\begin{abstract}

We report small angle neutron scattering (SANS) measurements of the skyrmion lattice in (Cu$_{0.976}$Zn$_{0.024}$)$_2$OSeO$_3$ under the application of an electric field. These measurements show an expansion of the skyrmion lattice stability region with electric field. Furthermore, using time-resolved SANS, we observe the slow formation of skyrmions after an electric or magnetic field is applied, which has not been observed in pristine Cu$_2$OSeO$_3$ crystals.  The measured formation times are dramatically longer than the corresponding skyrmion annihilation times after the external field is removed, and increase exponentially from 100~s at 52.5~K to 10,000~s at 51.5~K. This thermally activated behavior indicates an energy barrier for skyrmion formation of 1.57(2)~eV, the size of which demonstrates the huge cost for creating these complex chiral objects.

\end{abstract}

\maketitle

\section{Introduction}

Topological states of matter are the subject of extensive interest in condensed matter physics~\cite{Duncan2017,Hasan2010,Qi2012,Nagaosa2013}. One such state, the skyrmion lattice, consists of topologically protected nanoscale magnetic solitons that form a hexagonal lattice~\cite{Rossler2006}. Skyrmion lattices usually form only in a small region of magnetic field and temperature just below the magnetic ordering temperature, and are typically stabilized by competition between the Dzyaloshinskii-Moriya interaction (DMI), symmetric exchange, and thermal fluctuations~\cite{Nagaosa2013}. This state was first discovered in the non-centrosymmetric metallic compound MnSi which crystallizes in the $P2_13$ space group~\cite{Muhlbauer2009}, and has since been found in isostructural materials such as FeGe~\cite{Yu2011}, Fe$_{1-x}$Co$_x$Si~\cite{Munzer2010}, and Cu$_2$OSeO$_3$~\cite{Seki2012}, in other noncentrosymmetric compounds~\cite{Tokunga2015, Fujima2017,Ruff2015}, in thin films where skyrmions are stabilized by interfacial DMI~\cite{Heinze2011, Moreau2016, Anjan2017} or by a combination of DMI, uniaxial anisotropy and geometric confinement \cite{Leonov2016,Wilson2014,Rybakov2013}, and in some centrosymmetric materials where the state is thought to be stabilized by magnetic frustration~\cite{Kurumaji2018, Hirschberger2018}.

Skyrmions have been proposed for spintronics applications as stable, nanoscale, storage elements~\cite{Fert2013, Fert2017}. One key requirement for this application is the ability to reliably perform a write operation by nucleating a skyrmion~\cite{Romming2013}. Cu$_2$OSeO$_3$ is seen as attractive from this perspective as it is an insulating magnetoelectric whose skyrmion phase stability can be enhanced by applying electric fields~\cite{Kruchkov2018}. Furthermore, applying an electric field has been shown to nucleate skyrmions from either the competing conical state in bulk crystals ~\cite{White2018}, or the competing helical state in thin lamellae~\cite{Huang2018}. 

Another method for expanding the skyrmion region is to rapidly cool the system in a magnetic field to form metastable skyrmions. In Cu$_2$OSeO$_3$~\cite{Okamura2016} and other materials~\cite{Milde2013,Oike2016,Karube2016}, this produces skyrmions that, at low temperature, exist for much longer than typical observation times. However, near the border of the equilibrium skyrmion pocket, metastable skyrmions are observed to annihilate with lifetimes shorter than are experimentally accessible (seconds)~\cite{Oike2016}. As a result of this short lifetime at higher temperatures, rapid cooling is required to stabilize a sizable metastable skyrmion population.

Recently, crystals of (Cu$_{1-x}$Zn$_x$)$_2$OSeO$_3$ ($x$ = 0 to 0.024) have been grown and shown to host similar skyrmion phases to pristine Cu$_2$OSeO$_3$, but at temperatures shifted slightly lower as the Zn substitution level is increased~\cite{Stefancic2018}. Notably, the lifetime of metastable skyrmions in Zn substituted crystals is dramatically enhanced compared to that seen in pristine Cu$_2$OSeO$_3$. This suggests a high sensitivity of the energetic balance between the conical and skyrmion phases to Zn substitution \cite{Birch2018}, in addition to the known sensitivity to applied electric fields \cite{Kruchkov2018, White2018}. It is therefore an interesting open question to determine the interplay of these two effects by studying how electric fields affect the skyrmion phase in (Cu$_{1-x}$Zn$_x$)$_2$OSeO$_3$ in comparison to how they affect skyrmions in pristine Cu$_2$OSeO$_3$.

In this paper, we use small angle neutron scattering (SANS) to probe the skyrmion state in a crystal of (Cu$_{0.976}$Zn$_{0.024}$)$_2$OSeO$_3$. We show that applying a positive electric field ($\vec{E} \parallel \vec{H}$) expands the temperature extent of the skyrmion pocket, while a negative electric field  suppresses it, as is seen in pristine Cu$_2$OSeO$_3$ \cite{Kruchkov2018}. Furthermore, we find that skyrmions formed by the application of an electric field take a measurably long time to appear (100 - 10,000~s). The characteristic formation time is temperature dependent, allowing us to extract an energy barrier for the formation of skyrmions of 1.57(2)~eV. This time-dependent behavior has not been observed in any other bulk skyrmion material. Our measurement of the energy barrier for skyrmion formation is an important advance in the understanding of skyrmion behavior.

\begin{figure}[t]
\includegraphics[width=\columnwidth]{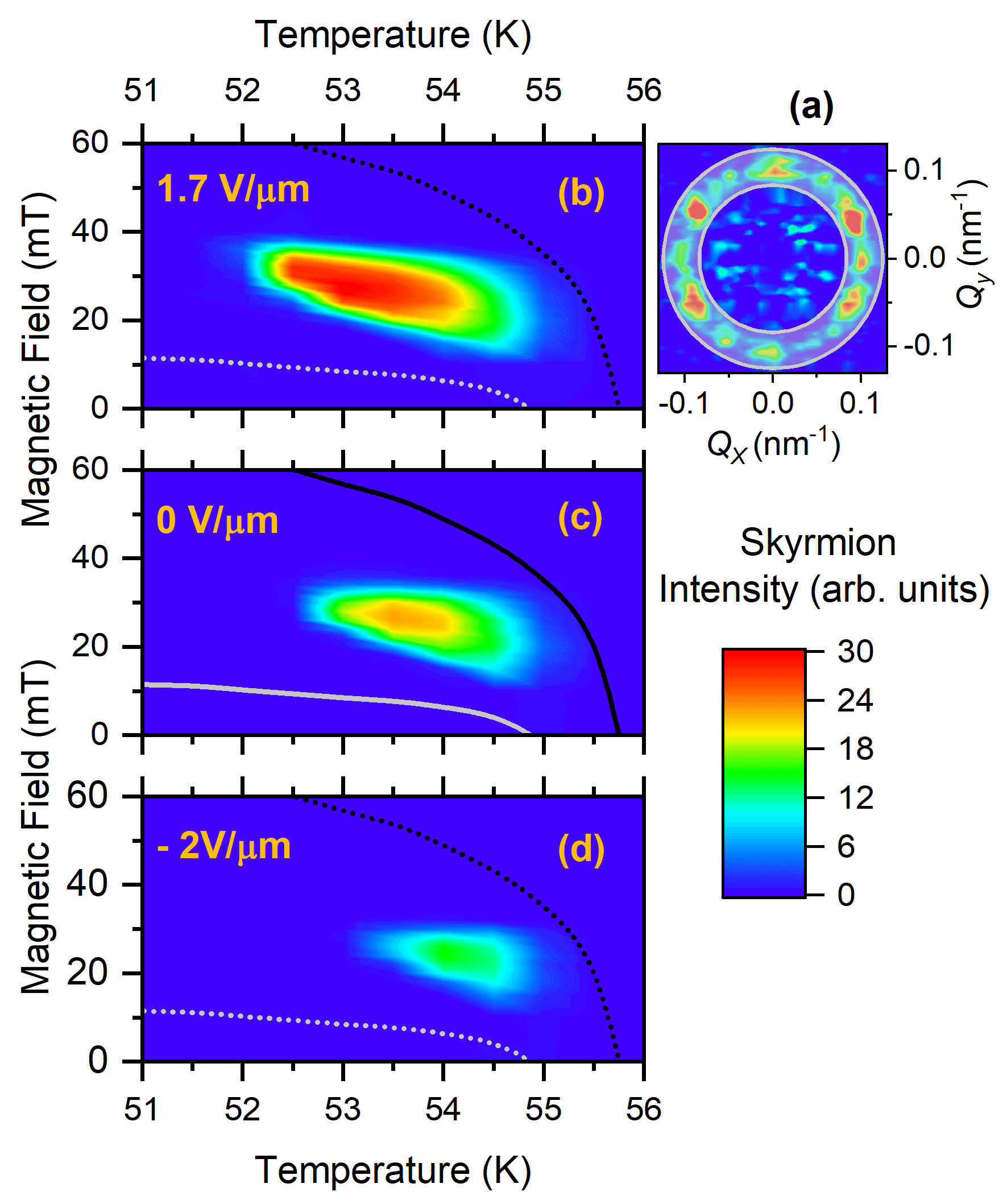}
\caption{(a) SANS pattern measured at 54 K with $E = -2$~ V/$\mu$m and $\mu_0H = 26$~mT. The gray ring shows the area that was summed on each pattern to characterize the skyrmion intensity. (b-d) Skyrmion phase diagrams of (Cu$_{0.976}$Zn$_{0.024}$)$_2$OSeO$_3$ measured by SANS with increasing magnetic field sweeps after zero magnetic field cooling in applied electric fields of (b) +1.7~V/$\mu$m, (c) 0, and (d) $-2$~V/$\mu$m. The lines show the conical - field polarized boundary (black) and the helical - conical boundary (gray) measured by AC susceptibility at E = 0. }
\label{fig:PDE}
\end{figure}
\section{Experiment Details}

A single crystal of (Cu$_{0.976}$Zn$_{0.024}$)$_2$OSeO$_3$ ($T_C$~=~55.8~K) was grown by chemical vapor transport (See Ref. \cite{Stefancic2018} for details). From this crystal, a 1.08 x 3 x 3 mm plate (thin axis $\parallel$ [111]) was cut and coated on both sides with silver paste to form conductive plates. With thin copper wires attached to the silver paste, the crystal was mounted on a sapphire plate in the Paul Scherrer Institut (PSI) high voltage sample stick \cite{Bartkowiak2014} and placed under vacuum. The evacuated sample stick was then inserted into a horizontal field cryomagnet on the PSI SANS-II instrument~\cite{Strunz2004} to perform SANS measurements with 10~\r{A} neutrons, under a magnetic and electric field $\vec{E} \parallel \vec{H} \parallel$ [111] crystal axis. SANS data analysis was performed using the GRAS$_{\text{ANS}}$P software (v8.11b) from the Institut Laue-Langevin~\cite{Dewhurst2003}. AC susceptibility measurements were performed with $\vec{H} \parallel$ [111] using a Quantum Design MPMS-5S with an AC driving field of 3 Oe.

\section{Results and Discussion}

Figure \ref{fig:PDE} (a) shows a typical skyrmion SANS pattern taken at 54 K with $E = -2 $~V/$\mu$m and $\mu_0H = 26$~mT. The six-fold scattering pattern is characteristic of a skyrmion lattice with 61.30(1)~nm spacing. Intensity in a ring between the main scattering spots arises from rotational disorder in the skyrmion lattice. To characterize the intensity of the skyrmion state (proportional to sample volume in the skyrmion state), we sum the intensity in the gray ring indicated on Fig. \ref{fig:PDE} (a).

Figures \ref{fig:PDE} (b - d) show the skyrmion intensity as a function of temperature and magnetic field for three different applied electric fields. The data show an enhancement and expansion of the skyrmion state by 0.5~K/(V/$\mu$m)  when a positive electric field is applied, and a corresponding  suppression of the skyrmion state when a negative electric field is applied, consistent with that seen in pristine Cu$_2$OSeO$_3$~\cite{Kruchkov2018}.

\begin{figure}[t]
\includegraphics[width=\columnwidth]{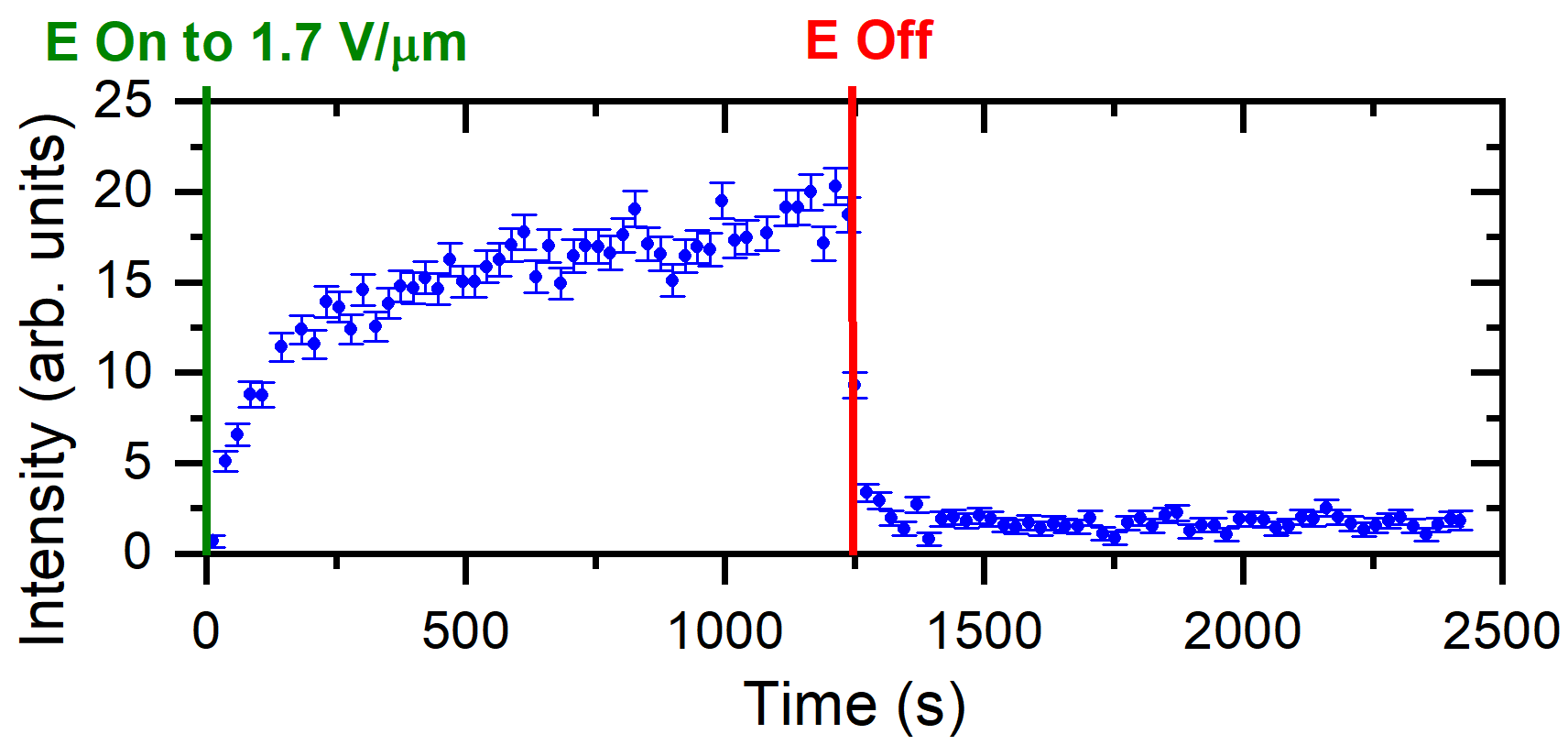}
\caption{Time dependence of skyrmion SANS intensity when applying and removing a 1.7 V/$\mu$m electric field at 52.25~K and 34~mT.}
\label{fig:NucAni}
\end{figure}

To further investigate the behavior of the skyrmions when an electric field is applied, we performed time-resolved SANS in the region $T \le 52.6$~K where the phase diagram indicates that skyrmions exist for an applied positive electric field but not for zero electric field. For these measurements we zero field cooled to the desired temperature, turned on the magnetic field, began ramping up the electric field, and commenced counting. After measuring the skyrmion lattice formation, we turned off the electric field, leaving the magnetic field on, and continued to count. A sample of these measurements in shown in Figure \ref{fig:NucAni} for $\mu_0H = 34 $~mT and $T = 52.25$~K.

This data shows that the skyrmion state takes significant time to form when the electric field is applied ($\tau_{\text{form}} \approx 200$~s), but disappears much more rapidly when the electric field is removed ($\tau_{\text{annihilate}} < 20$~s). Such behavior suggests a much larger energy barrier to form a skyrmion than to annihilate one. Further, Fig. \ref{fig:NucAni} shows that the skyrmion intensity does not return to zero after the electric field is removed, suggesting that the equilibrium skyrmion region at zero electric field extends below the phase diagram in Fig. \ref{fig:PDE} (c). This likely indicates that at low temperature the formation time becomes long compared to the measurement timescale. As the annihilation time is very short, we could not accurately measure it, and we therefore focus on the formation time for the remainder of this paper.

\begin{figure}[t]
\includegraphics[width=\columnwidth]{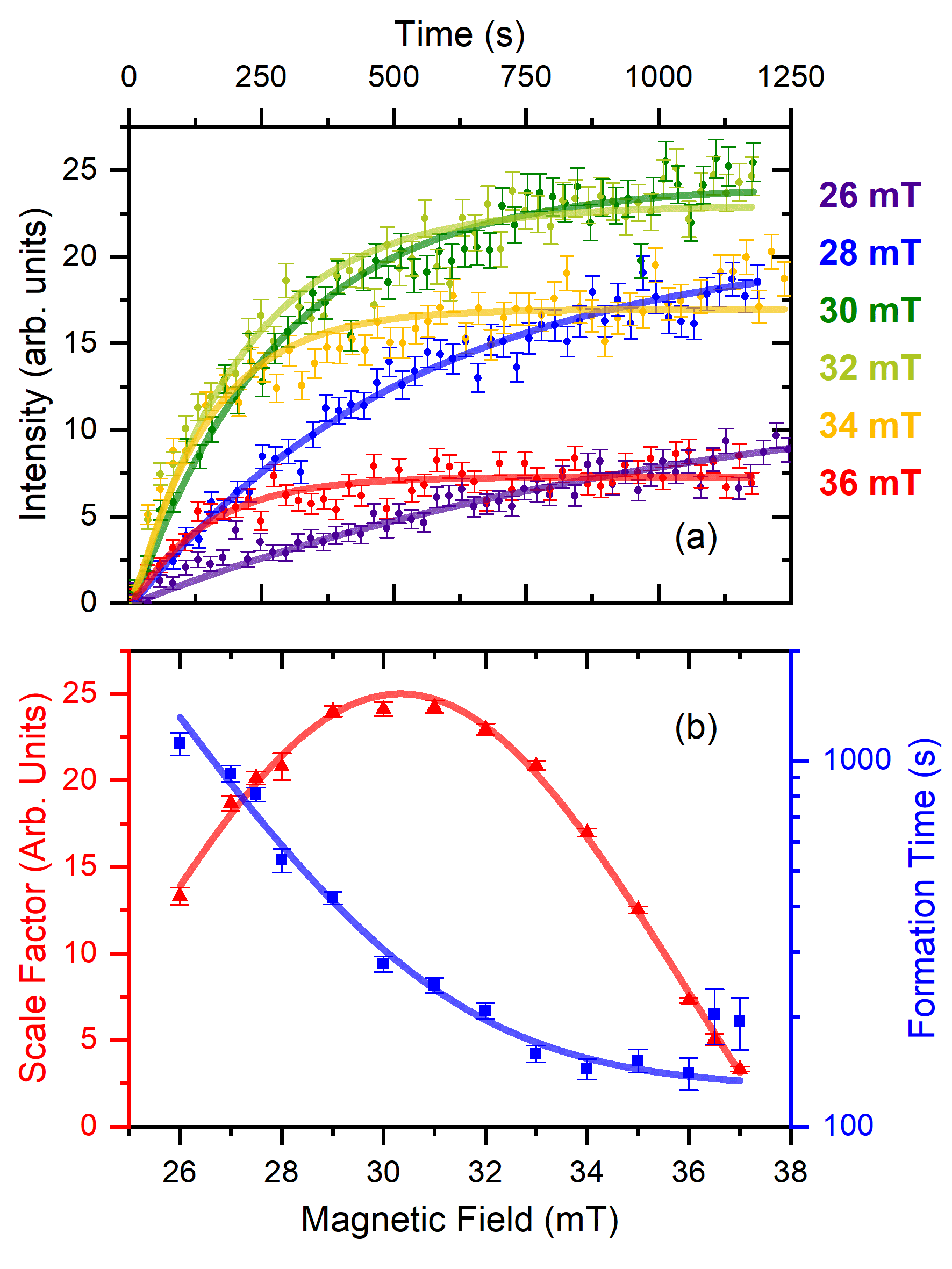}
\caption{Formation of skyrmions by applying an electric field of 1.7~V/$\mu$m at 52.25~K in varied magnetic fields. (a) Skyrmion SANS intensity as a function of time. (b) Fit formation time (blue squares) and intensity scale factor (red triangles) from Eq. \ref{eq:nucleate}. Solid lines in (b) show fits to a Gaussian peak and $\tau = A e^{-\lambda \mu_0H} + \tau_a$.}
\label{fig:TauB}
\end{figure}

To fit the skyrmion intensity $S$ accounting for the ramping of the electric field (See Appendix), we use the equations,

\begin{equation}
S = 
\begin{cases}
    A\left[1 - (1+kt)^{\frac{-1}{ka}}\right]&  t\leq t_0\\
    \ \\
    A\left[1 - e^{\frac{-t}{\tau}} \frac{ (1+kt_0)^{\frac{-1}{ka}}}{e^{\frac{-t_0}{\tau}}}\right] &  t > t_0 .
\end{cases}
\label{eq:nucleate}
\end{equation}

\begin{figure}[ht!]
\includegraphics[width=\columnwidth]{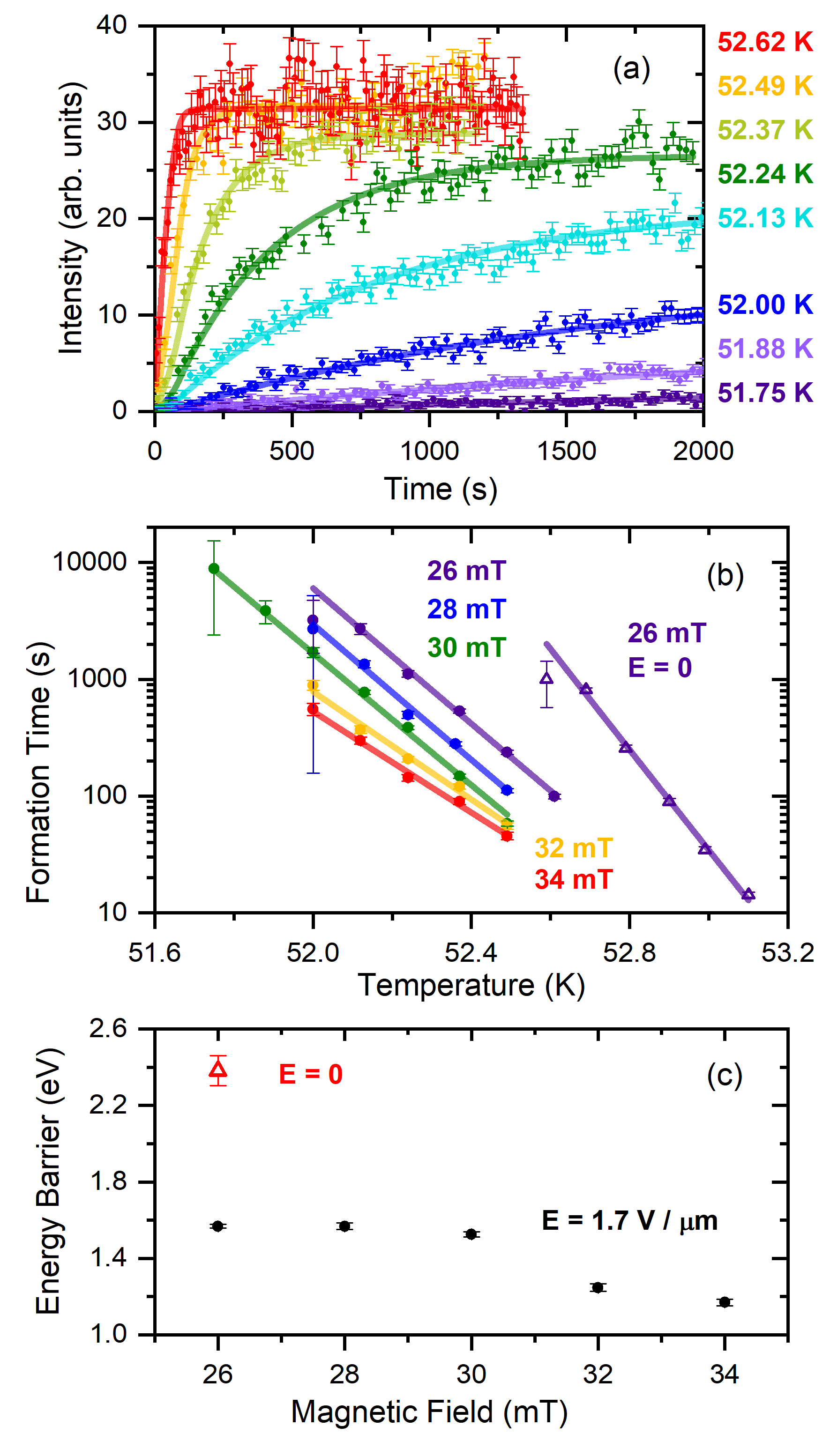}
\caption{(a) Skyrmion SANS intensity as a function of time after switching on $E = 1.7$ V/$\mu$m at $\mu_0H = 30$ mT for temperatures between 51.75 and 52.62 K. (b) Formation times as a function of temperature. Circles show formation times after switching on $E = 1.7$ V/$\mu$m in various magnetic fields. Purple triangles show the formation time after switching on $\mu_0H$ = 26 mT in zero applied electric field. Lines show fits to an Arrhenius law. (c) Energy barriers from fits in panel (b) for $E = 1.7$ V/$\mu$m (black circles) and $E = 0$ (red triangle). }
\label{fig:TauT}
\end{figure}

Here, $A$ is a scale factor for the skyrmion intensity, $\tau = a(1+kt_0)$ is the reported formation time, $a$ is allowed to vary with temperature and magnetic field, k accounts for the electric field ramp rate ($k = -0.0133$ s$^{-1}$ for Fig. \ref{fig:TauB} and $-0.008$ s$^{-1}$ for Fig. \ref{fig:TauT}), and $t_0$ is the time when the electric field finished ramping.

Using this fitting function, we characterized the skyrmion formation time across the magnetic field - temperature phase diagram. Figure \ref{fig:TauB} (a) shows measurements of skyrmion formation after applying an electric field of 1.7 V/$\mu$m at $52.25$~K, as a function of magnetic field. Between each measurement, we reset the magnetic state by setting both the magnetic and electric field to zero at 52.25~K. Figure \ref{fig:TauB} (b) shows the fitted scale factor and formation time as a function of field. The scale factor is well described by a Gaussian curve centered at 30.3(1)~mT. The formation time is asymmetric across the skyrmion pocket, at first rapidly decreasing, then asymptoting to $\tau_a = 130(10)$~s at high field, phenomenologically fitting to $\tau = A e^{-\lambda \mu_0H} + \tau_a$, where $\lambda = 0.48(5)$~mT$^{-1}$.

 Figure \ref{fig:TauT} (a) shows skyrmion formation after applying a 1.7 V/$\mu$m electric field for various temperatures in an applied magnetic field of 30~mT. These data indicate an increasing formation time for decreasing temperature, characteristic of thermal activation over an energy barrier. In Fig. \ref{fig:TauT} (b), we show fits of the formation times to an Arrhenius law, $\tau = \tau_0 \text{e}^{\frac{E}{k_{\text{B}} T}}$, to extract the energy barrier $E$, where $\tau_0$ is a scale factor, and $T$ is temperature. Fits of data obtained at five different applied magnetic fields yield energy barriers between 1.57(2)~eV (26~mT) and 1.17(2)~eV (34 mT), as shown by the black circles in Fig. \ref{fig:TauT}.
The energy barrier is nearly constant up to 30 mT, decreasing thereafter for higher magnetic fields. This change in the energy barrier is consistent with the magnetic field values where the formation time in Fig. \ref{fig:TauB} (b) deviates from exponential behavior, suggesting that the reason for the apparent saturation of the formation time at high fields is a change in the skyrmion formation energy barrier.

\begin{figure}[ht!]
\includegraphics[width=\columnwidth]{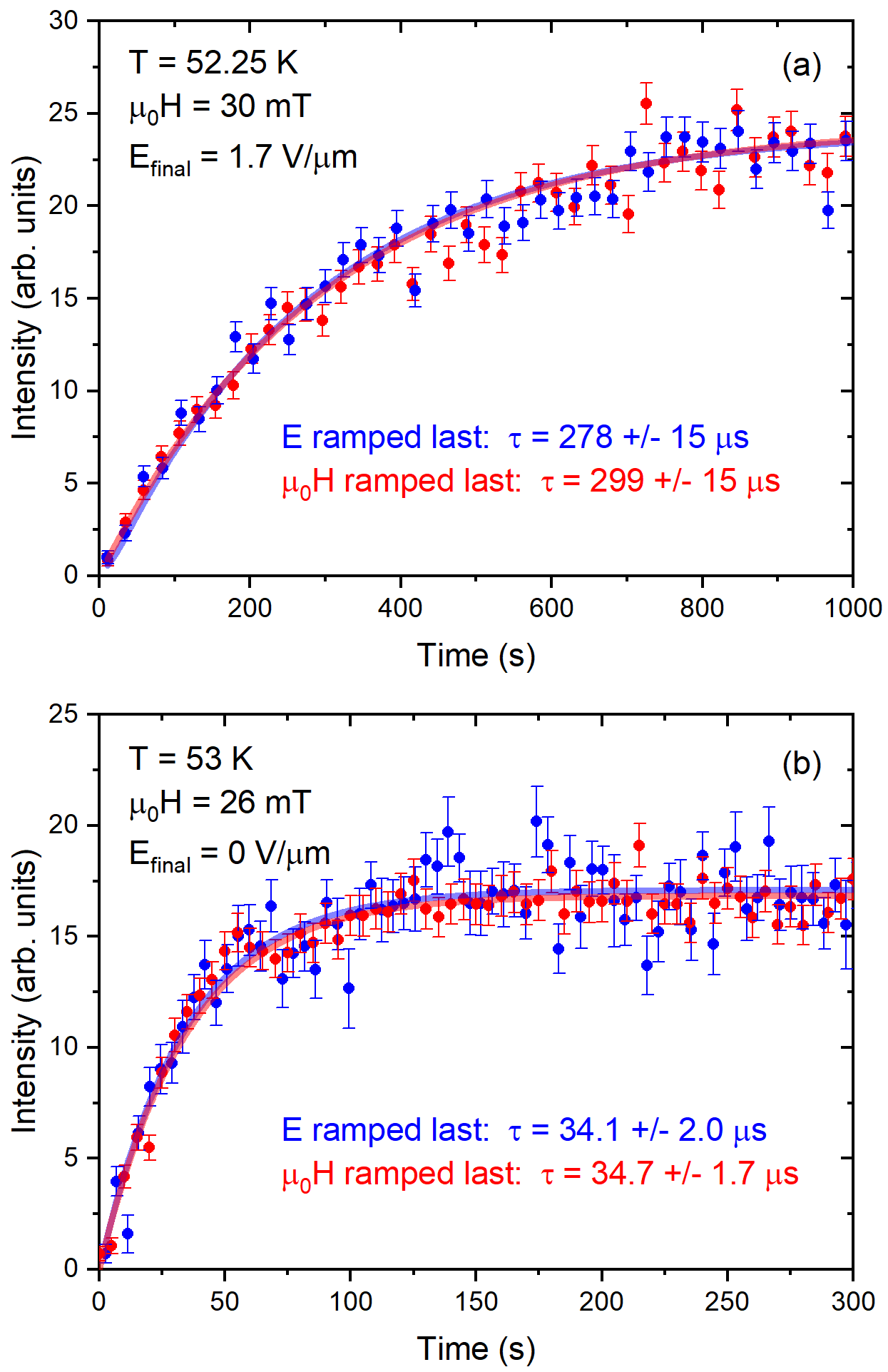}
\caption{ (a) Skyrmion SANS intensity as a function of time measured at $E = 1.7$ V/$\mu$m, $\mu_0H = 30$ mT, and T = 52.25~K. The red data show measurements when the electric field was turned on first, and then the magnetic field was turned on at $t=0$~s. These are fit to $S = A(1 - e^{-\frac{t}{\tau}})$. The blue data show measurements when the magnetic field was turned on first, and the electric field ramping begins at $t=0$~s. These are fit to Eq. \ref{eq:nucleate}. (b) Skyrmion SANS intensity as a function of time measured at $E = 0$ V/$\mu$m, $\mu_0H = 26$ mT, and T = 53~K. The red data show measurements where the magnetic field was turned on at time = 0 in zero electric field. The blue data show measurements in a constant magnetic field where the electric field was changed from $-1.7$~V/$\mu$m to 0 at $t=0$~s. Both data sets were fit to $S = A(1 - e^{-\frac{t}{\tau}})$ as the magnetic field changes quickly and the electric field changes quickly when set to zero.}
\label{fig:EvB}
\end{figure}

 To determine whether this skyrmion formation behavior is solely an electric field effect, or inherent to the skyrmion state, we performed time-dependent measurements when applying a magnetic field in constant electric field to compare with the process of turning on an electric field in a constant magnetic field. As our cryomagnet changed the magnetic field very rapidly, $S = A(1 - e^{-\frac{t}{\tau}})$ fits this data with no correction required for ramping time. These data are shown in Fig. \ref{fig:EvB}~(a) and demonstrate that the behaviour is the same for both processes, yielding formation times that agree within their uncertainties. As an additional check, we compared measurements performed after switching a magnetic field on in zero electric field, to measurements performed in a constant magnetic field after switching an electric field of $-1.7$~V/$\mu$m (which suppresses the skyrmion stability region) to zero (where skyrmions form). Shown in Fig. \ref{fig:EvB}~(b), these data again indicate no difference between the skyrmion formation behaviour observed after the two processes. Figure \ref{fig:EvB} therefore shows that the factor affecting the formation time is the final magnetoelectric field state of the system, not the process by which it was brought there. 

For comparison with the behavior observed in an applied electric field, we measured the temperature dependence of formation times when switching on a 26~mT magnetic field in zero electric field. Shown in Fig. \ref{fig:TauT} (b), these formation times display a temperature dependence consistent with thermally activated behavior, similar to the data measured in a 1.7~V/$\mu$m electric field. However, for the same magnetic field, the 2.38(8) eV energy barrier measured with zero electric field (red triangle in Fig. \ref{fig:TauT} (c)) is substantially larger than the 1.57(2) eV barrier measured with an applied electric field, demonstrating that the electric field suppresses the energy barrier between the conical and skyrmion states.

Energy barriers for skyrmion annihilation have previously been measured in systems such as FeGe (0.19 eV) \cite{Peng2018}, Fe$_{0.5}$Co$_{0.5}$Si (0 - 0.2 eV) \cite{Wild2017}, and Cu$_2$OSeO$_3$ (0.4~eV) \cite{Birch2018}, and are substantially smaller than our measured energy barriers for skyrmion formation. This is consistent with Fig.~\ref{fig:NucAni} showing that skyrmion annihilation is faster than skyrmion formation, and with theoretical predictions of smaller energy barriers for skyrmion annihilation than for skyrmion formation\cite{Buttner2018}. 
 
Formation times long enough to measure with SANS have not previously been seen in similar electric-field switching experiments on crystals of pristine Cu$_2$OSeO$_3$ \cite{White2018}, which suggests that the formation times in our Zn-substituted samples are dramatically longer than those in the pristine material. Previous measurements of Zn substituted crystals suggest that the lifetime of metastable skyrmions is substantially enhanced by entropic effects arising from the non-magnetic substitution~\cite{Birch2018}, which may also explain the long formation times. However, it is also possible that Zn substitution does not change the skyrmion formation physics, and it is simply a reduction of the lower boundary of the skyrmion region from 55.5~K in pristine Cu$_2$OSeO$_3$ to 52.5~K in our Zn substituted sample~\cite{Stefancic2018} that lengthens the formation time. Extrapolating our zero electric field data to 55.5~K, we would expect a formation time of $2.1^{+4.6}_{-0.9}$~ns, which is too short to observe with SANS.

It has previously been proposed that skyrmions form~\cite{Sampaio2013} and are annihilated~\cite{Milde2013, Kagawa2017} by the nucleation of Bloch points to create skyrmion strings, which then grow by propagation of the Bloch points. SANS measurements are sensitive to the average length of skyrmion strings by the width of rocking curves obtained by measuring the SANS intensity while rotating the sample and field together around the incoming beam direction (angle $\alpha$). If the skyrmion strings are short, the rocking curve width will be large, and as the strings grow the width is expected to decrease. 

\begin{figure}[t]
\includegraphics[width=\columnwidth]{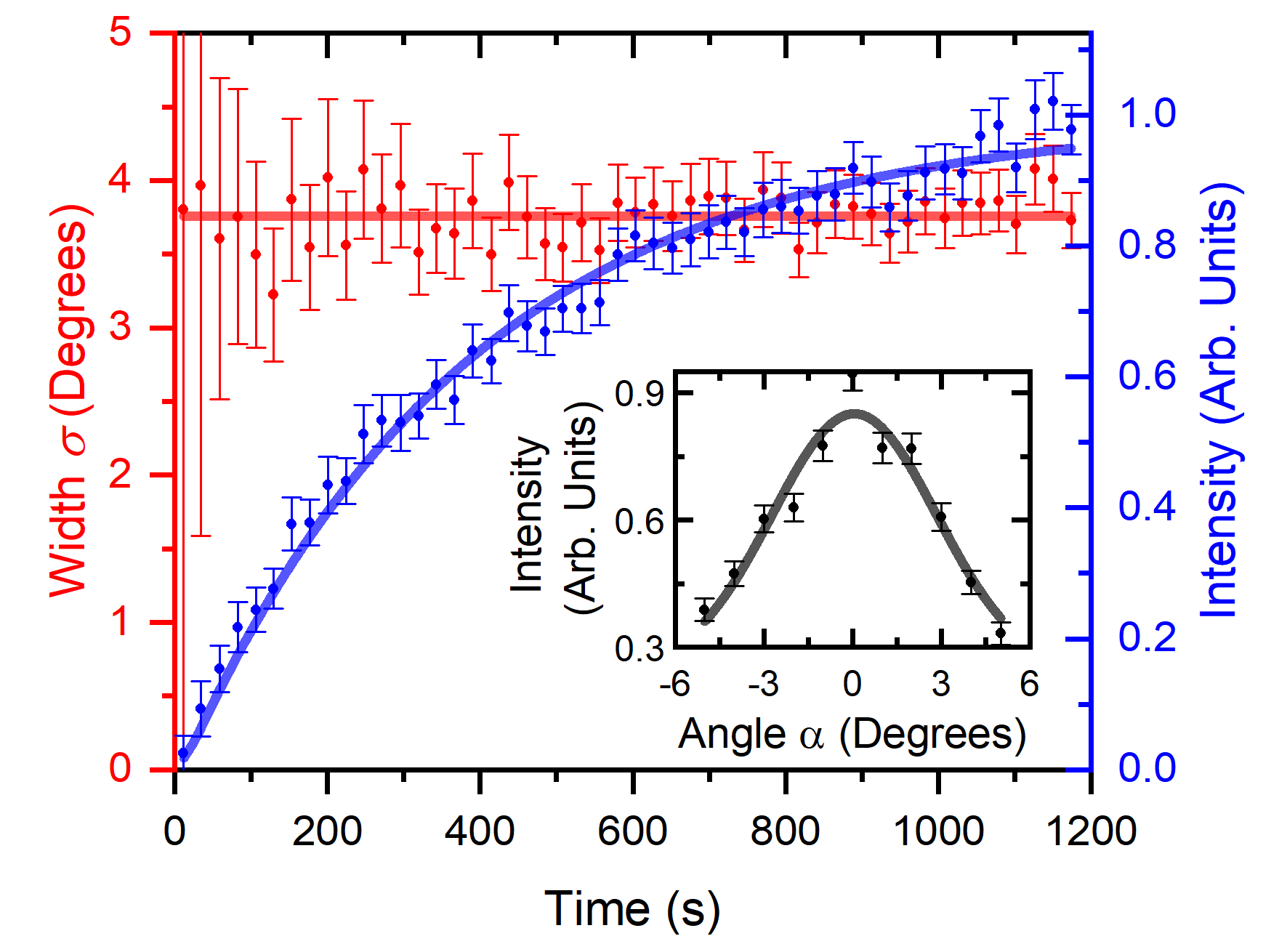}
\caption{Gaussian widths and intensities for SANS rocking curves measured during skyrmion formation at 52.25 K in a 30~mT magnetic field after applying a 1.7 V/$\mu$m electric field. The inset shows an example rocking curve constructed from data taken 20 minutes after applying the electric field.}
\label{fig:Rock}
\end{figure}

To study this process, we performed a series of skyrmion formation measurements at different angles ($-5.5$\textdegree$ \le \alpha \le 5.5$\textdegree), and stacked them to form effective rocking curves as a function of time, an example of which is shown in the inset of Fig. \ref{fig:Rock}. The fit Gaussian widths of these curves are displayed in Fig. \ref{fig:Rock} and show no change as a function of time. This suggests that the skyrmion length does not slowly increase over time, but rather that skyrmions either initially appear as long strings, or that after nucleating small skyrmions, Bloch points propagate too fast to be observed with SANS. The increasing SANS intensity therefore most likely arises from an increasing number of skyrmions: Bloch point nucleation, rather than propagation, is the rate-limiting step for skyrmion formation. However, we note that our minimum rocking curve width (corresponding to a skyrmion length of 150~nm) is inherently broadened by instrumental resolution and variations in skyrmion string alignment, reducing our sensitivity to long length scales.

As we infer that the SANS intensity increase corresponds to an increase in the number of skyrmions, our measured energy barrier is that required to nucleate a Bloch point and start formation of a skyrmion string. The energy barrier of 1.5 - 2.5 eV is enormous compared to the thermal energy of single spins (400~k$_{\text{B}}T$ for $T$ = 52~K), suggesting that Bloch points are extended multi-spin objects rather than single spins. In Cu$_2$OSeO$_3$ the base spin units are ferrimagnetic tetrahedra with three parallel and one antiparallel Cu$^{\text{2+}}$ spins~\cite{Romhanyi2014,  Franke2018}. Considering the exchange interactions of $J = -3.6$~meV~\cite{Tucker2016}, the energy to flip a single Cu tetrahedron spin unit is $\approx$ 0.1~eV. This suggests that nucleating a Bloch point requires flipping $\approx 20$ tetrahedra, corresponding to an initial size of a couple nanometers, which is far smaller than the 61~nm skyrmion diameter. Our data therefore indicate that while Bloch points in chiral magnets are small relative to skyrmions, they are significantly larger than single spins.

\section{Conclusions}

In conclusion, we have measured the temperature and magnetic field dependence of the skyrmion formation time in Zn substituted Cu$_2$OSeO$_3$. This shows thermally activated behavior, with time scales ranging from 10,000 seconds at the lowest temperatures down to 100 seconds at the highest temperatures, giving a characteristic energy barrier for skyrmion formation of 1.57(2) eV. Combining our results for skyrmion formation with previous measurements of skyrmion annihilation allows a more complete picture of the energetics of skyrmions to be inferred, which will be highly important for future research towards skyrmionic devices.

\begin{acknowledgements}
We thank T.J. Hicken and T. Lancaster for useful discussions, as well as J. Debray and M. Bartkowiak for technical support. The neutron scattering experiments were performed at the Swiss Spallation Neutron Source (SINQ), Paul Scherrer Institut, Switzerland. This work was financially supported by the UK Skyrmion Project EPSRC Programme Grant (EP/N032128/1) and the Swiss National Science Foundation (SNF) Sinergia network ``NanoSkyrmionics'' grant (CRSII5-171003). M.~N.~Wilson acknowledges the support of the Natural Sciences and Engineering Research Council of Canada (NSERC). 
\end{acknowledgements}

\appendix*
\section{Electric Field Dependence}

\renewcommand{\thefigure}{A\arabic{figure}}
\setcounter{figure}{0} 

\begin{figure}[h]
\includegraphics[width=\columnwidth]{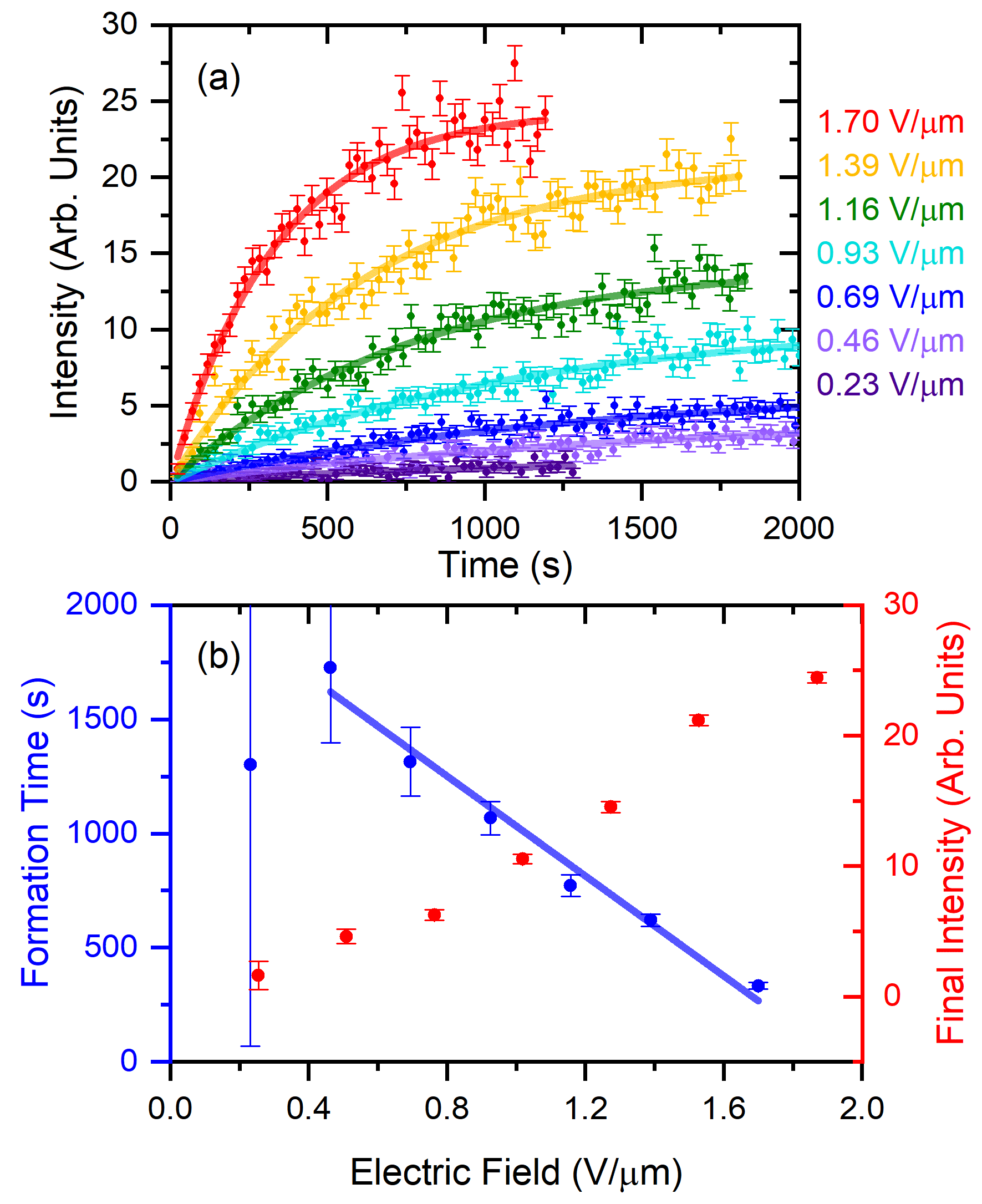}
\caption{Formation of skyrmions at 52.25 K in a magnetic field of 30 mT after applying various electric fields.}
\label{fig:TauV}
\end{figure}

For each of the measurements of the skyrmion formation time in an electric field, the ramping time of the electric field was non-negligible compared to the total measurement time. The electric field was ramped quickly from 0 to 0.9 V/$\mu$m (Fig. 3)  or 0 to 1.4 V/$\mu$m (Fig. 4), and then at 0.009 (V/$\mu$m)/s up to the peak electric field of 1.7 V/$\mu$m, giving a ramp time ($t_0$) of 33.6 or 83.6 s. The formation time of skyrmions depends linearly on the electric field, as shown by Fig. \ref{fig:TauV}, fitting to $\tau(E) = m E + b$. The ramp time for the electric field therefore requires a small correction to the fitting procedure to produce accurate formation times.

We first start with an exponential fitting function for the skyrmion intensity ($S$),

\begin{equation} 
S = A(1 - e^{-\frac{t}{\tau}}),
\label{eq:expAp}
\end{equation}
where $A$ is a scale factor representing the final intensity at long times, $t$ is the time, and $\tau$ is the formation time.

Next, assume $\tau$ varies linearly with time while the electric field changes, $\tau = a(1+kt)$ for  $t \le t_0$, where $t_0$ is the end time of the electric field ramp, $a$ is a scale factor, and $k$ is the relative slope which we assume to be fixed as a function of magnetic field and temperature. Defining $E_i$ as the electric field at the start of the slow ramp, and $\epsilon$ as the ramp rate, $k$ can be determined from the fit in Fig. \ref{fig:TauV} (b) as $k = \epsilon m / (b + m E_i) = 0.008$~s$^{-1}$ for $E_i$ = 0.9~V/$\mu$m and 0.0133~s$^{-1}$ for $E_i$ = 1.4~V/$\mu$m.

Now, writing $S(t) = A[1 - C(t)]$, the differential equation appropiate for an exponential process $C$ is,

\begin{equation}
dC/dt = -\frac{C}{\tau}. 
\end{equation}

Allowing $\tau$ to depend on time as $\tau = a(1+kt)$, this gives

\begin{equation}
dC/dt = -\frac{C}{a(1+kt)}.
\end{equation}

Solving this equation yields,

\begin{equation}
C = (1+kt)^{\frac{-1}{ka}}.
\end{equation}

Hence, the skyrmion intensity for $t \le t_0$ is given by,

\begin{equation}
S = A\left[1 - (1+kt)^{\frac{-1}{ka}}\right].
\label{eq:PL}
\end{equation}

After $t_0$ the skyrmion intensity will be given by the exponential relationship (Eq. \ref{eq:expAp}), scaled appropriately such that it is equal to Eq. \ref{eq:PL} at $t = t_0$. Therefore, the final expression for the skyrmion intensity at any time is,

\begin{equation}
S = 
\begin{cases}
    A\left[1 - (1+kt)^{\frac{-1}{ka}}\right]& \text{if } t\leq t_0\\
    \ \\
    A\left[1 - e^{\frac{-t}{\tau}} \frac{ (1+kt_0)^{\frac{-1}{ka}}}{e^{\frac{-t_0}{\tau}}}\right] & \text{if~}  t > t_0.
\end{cases}
\label{eq:nucleateAp}
\end{equation}

\begin{figure}[ht]
\includegraphics[width=\columnwidth]{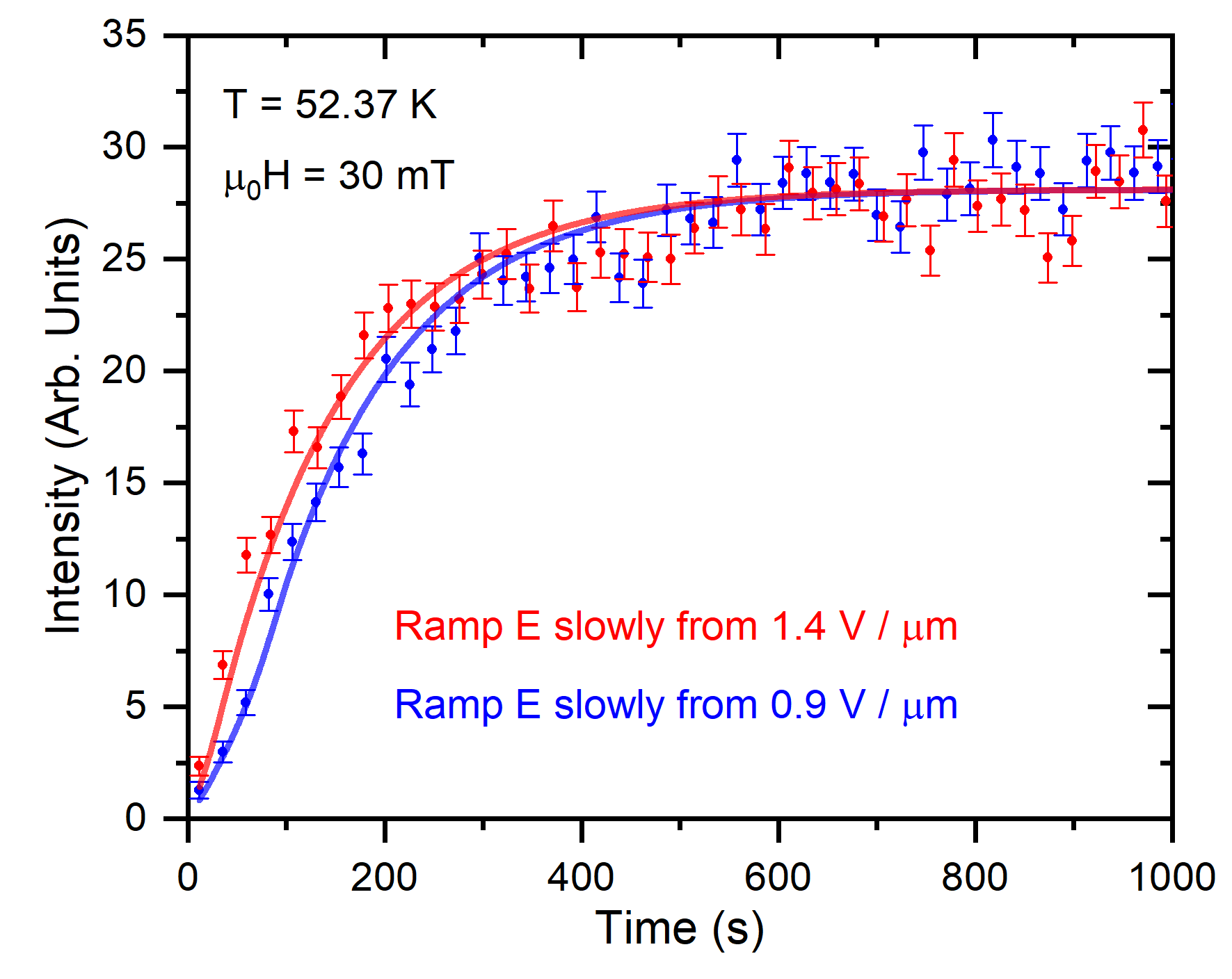}
\caption{Intensity of skyrmions as a function of time measured at T = 52.25~K and H = 30~mT for two different ramp rates of the electric field. Blue is for ramping fast to 0.9 V/$\mu$m and 0.009 (V/$\mu$m)/s up to 1.7 ~V/$\mu$m, while red is ramping fast to 1.4 V/$\mu$m and 0.009 (V/$\mu$m)/s up to 1.7 V/$\mu$m.  Solid lines show fits to Eq. \ref{eq:nucleateAp} using parameters for the two different ramp rates. Both are fit to the same formation time of 133 $\pm$ 5~s }
\label{fig:RampRates}
\end{figure}

This equation allows data at different electric field ramp rates to be fit with the same formation time ($\tau$), as shown by Fig. \ref{fig:RampRates}. In the paper, the values we report as the formation time are the constant values $\tau$ takes for $t > t_0$, $\tau = a(1+kt_0)$.

\end{document}